\documentclass[journal=esthag,manuscript=article]{achemso}
\usepackage[version=3]{mhchem} % Formula subscripts using \ce{}
\usepackage{graphicx} % Required for inserting images
\usepackage{xcolor}
\usepackage{array}
\usepackage{caption}
\usepackage{subcaption}

\usepackage{multirow}
\usepackage{makecell}

\usepackage{comment}
\usepackage{xr}
\usepackage{amssymb}
\usepackage{float}
\usepackage{pdfpages}

\author{Audrey Ngambia}
\affiliation[University of Edinburgh, School of Chemistry]
{School of Chemistry, University of Edinburgh, Joseph Black Building, David Brewster Road, EH9 3FJ, United Kingdom}
\author{Anastasiia Gavrilova}
\affiliation [University of Edinburgh, School of GeoSci]
{School of GeoScience, University of Edinburgh, Alexander Crum Brown Road, EH9 3FF, United Kingdom}

\author{Haitao Huang}
\affiliation[University of Edinburgh, School of Chemistry]
{School of Chemistry, University of Edinburgh, Joseph Black Building, David Brewster Road, EH9 3FJ, United Kingdom}

\author{Zhuodong Lyu}
\affiliation[University of Edinburgh, School of Chemistry]
{School of Chemistry, University of Edinburgh, Joseph Black Building, David Brewster Road, EH9 3FJ, United Kingdom}

\author{Ondřej Mašek}
\affiliation [University of Edinburgh, School of GeoSci, UKBRC]
{UK Biochar Research Center, School of GeoScience, University of Edinburgh, Alexander Crum Brown Road, EH9 3FF, United Kingdom}
\author{Margaret Graham}
\email{margaret.graham@ed.ac.uk}
\affiliation [University of Edinburgh, School of GeoSci]
{School of GeoScience, University of Edinburgh, Alexander Crum Brown Road, EH9 3FF, United Kingdom}

\author{Valentina Erastova}
\email{valentina.erastova@ed.ac.uk}
\affiliation[University of Edinburgh, School of Chemistry]
{School of Chemistry, University of Edinburgh, Joseph Black Building, David Brewster Road, EH9 3FJ, United Kingdom}
\title[Biochar-Manganese Adsorption Studies]
{Decoupling Precipitation and Surface Complexation during Mn(II) Removal by Biochar via Experiments and Atomistic Simulations}

\begin{document}

\begin{tocentry}
  \centering
  \includegraphics[width=1\linewidth]{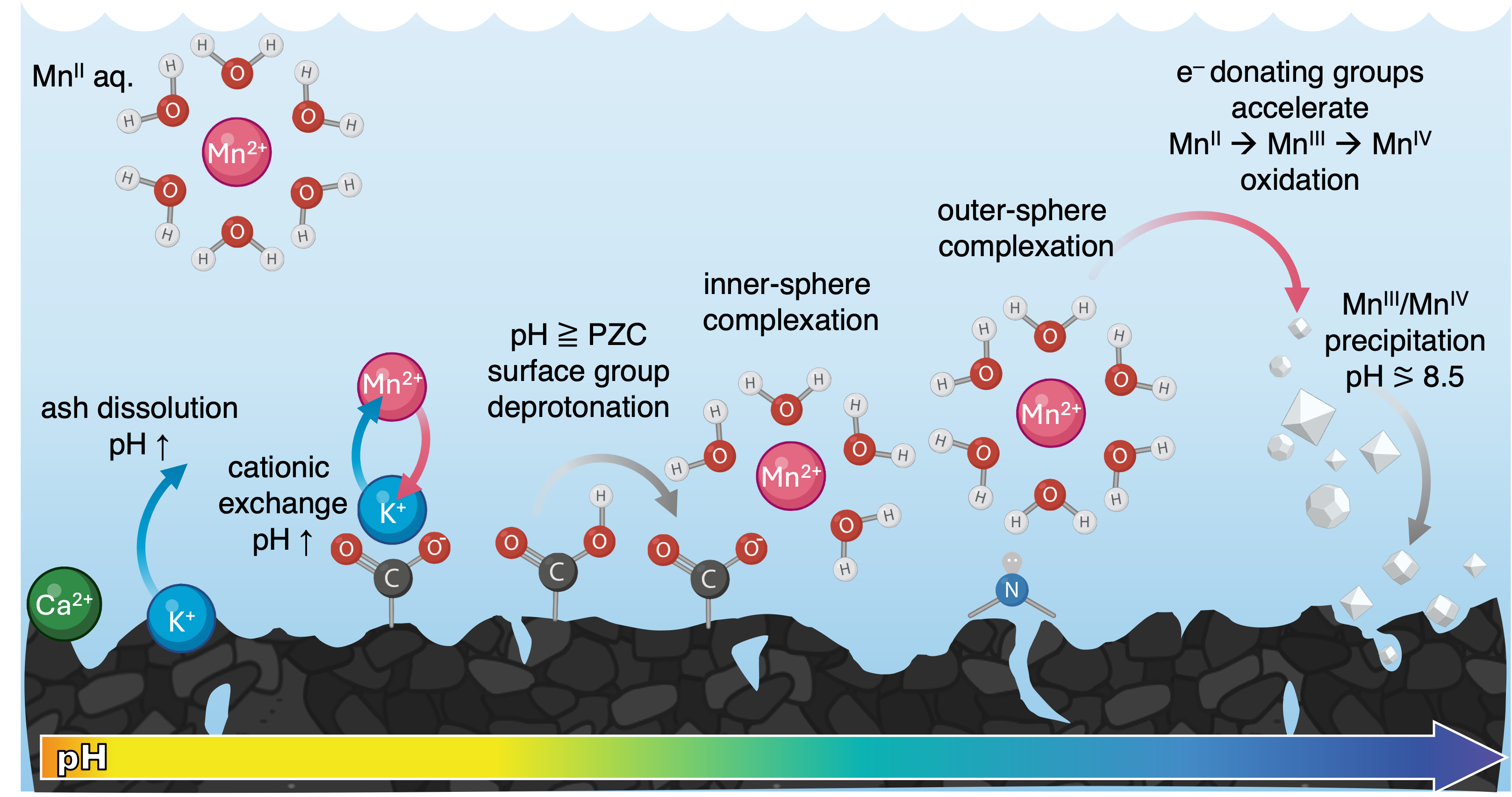}
\end{tocentry}


\begin{abstract}

Manganese(II) mobilised by mining activity poses a persistent water-quality challenge, yet the mechanisms by which low-cost sorbents, such as biochar, sequester Mn(II) remain poorly resolved. 
This study identifies the specific chemical drivers of Mn(II) sequestration by combining fixed-bed column and batch experiments with atomistic molecular dynamics simulations. 
Oilseed rape straw biochars, produced at 350\textdegree C, 550\textdegree C, and 700\textdegree C, removed 20-50\% of dissolved Mn from acidic influent (pH 4, 5 ppm). High-temperature biochar achieved the greatest removal ($\sim$50\%) and rapidly increased effluent pH to 9, triggering alkaline precipitation. 
Conversely, lower-temperature biochars removed 20-30\% of Mn while maintaining a near-neutral pH (7-7.5). Enhanced \ce{K+} release in these systems indicates significant cation exchange and non-precipitative pathways.
Molecular simulations confirmed that while neutral surfaces show weak Mn(II) association, deprotonated sites drive strong adsorption through inner-sphere complexation ($\sim$50\% removal) and outer-sphere association ($\sim$10\%). 
These results establish a mechanistic framework to distinguish between precipitation-led and surface-complexation-led removal. 
By providing specific chemical criteria for Mn-targeted sequestration, this work enables the rational design of engineered biochars for sustainable water remediation.

\end{abstract}

%%%%%%%%%%%%%%%%%%%%%%%%%%%%%%%%%%%%%%%%%%%%%%%%%%%%%%%%%%%%%%%%%%%%%
%% SYNOPSIS: 30 words non-tech statement articulating env context and impact
%%%%%%%%%%%%%%%%%%%%%%%%%%%%%%%%%%%%%%%%%%%%%%%%%%%%%%%%%%%%%%%%%%%%%

\textbf{Synopsis}:
Experiments and atomistic simulations decouple Mn(II) removal by biochar into exchange-driven pH increase, deprotonation-enabled surface complexation, and high-pH precipitation -- highlighting presence of deprotonatable functionalities as key design rules.

% The introduction should clearly and concisely explain the motivation for the work, its importance and originality, where it fits in the development of the field and why it should be of interest to ES&T readers. Discuss relationships of the study to previously published work, but do not reiterate or provide a complete literature survey. Current findings should not be included or summarized in this section. Introduction sections are typically around 500 words in length.

\section{Introduction}

Manganese (Mn) is a redox-active trace metal that becomes a water-quality concern when mobilised into surface and ground waters at elevated concentrations.\cite{cannon2014manganese,coles2012toxicological} 
While Mn is essential for biological function, sustained exposure to elevated dissolved Mn can pose ecological and human-health risks, and is increasingly recognised as an important contaminant in mining-impacted catchments.\cite{coles2012toxicological}
Mining activity is a major contributor, as extraction, processing, and mine-waste releases can mobilise Mn and co-occurring metals into adjacent waters. The Fund\~ao tailings dam failure in Mariana (Brazil, 2015) illustrates the persistence of Mn contamination following such events, with reported exceedances above regulatory thresholds across the affected watershed.\cite{fernandes2016deep,queiroz2018samarco,queiroz2021manganese}

Current treatment options for heavy metal contaminated waters (e.g., chemical precipitation, membranes, and biological approaches) can be costly, energy-intensive, or slow-acting, limiting their scalability. \cite{neculita2019review,chang2022effects,wu2022manganese} 
Adsorption offers a technically straightforward alternative, motivating the development of low-cost sorbents. \cite{tan2022enhancing}
Biochar is an attractive candidate because its carbon-rich structure contains surface functionalities capable of metal binding, while its mineral fraction can drive ion exchange and alkalinity changes.\cite{meng2025advances} 
However, these contributions often occur simultaneously, complicating mechanistic attribution and limiting the rational design of Mn-targeted biochars.

Indeed, several studies report substantial Mn removal by ash-rich biochars, yet most control only the initial solution pH and do not quantify pH evolution during exposure.\cite{idrees2018adsorption,kim2020removal,ma2023effect} This is important because dissolution of ash-derived alkali and alkaline-earth cations can raise solution pH into a regime where Mn(II) oxidation and precipitation of Mn(III/IV) (hydr)oxides becomes favourable.\cite{morgan1964colloid,gabelich2006sequential} 
(A more detailed discussion of prior Mn--biochar studies and the role of ash-driven pH evolution is provided in Section \ref{SI:Mn_literature}, SI.)
Consequently, apparent ``adsorption" capacities may reflect a coupled sequence of ion exchange, alkalinity generation, surface complexation, and precipitation rather than a single dominant pathway.

A recent study of the Standard Biochars further indicates strong feedstock dependence, with straw-derived biochars consistently outperforming woody biochars for Mn removal.\cite{roberta2026,mavsek2018consistency} 
Nevertheless, it remains unclear whether this behaviour is governed primarily by mineral inventory (ash/cationic exchange driven alkalinity increase and Mn oxidation and precipitation) or by the biochar-surface chemistry (functional-group density and speciation) that may enable Mn surface complexation.

Experimentally, it is difficult to disentangle the respective contributions of the biochar matrix, surface functional groups, ash-driven alkalinity, and exchangeable cations to Mn removal, because these processes co-occur and mutually reinforce one another during exposure. 
Therefore, we leverage our recent advances in experimentally-constrained atomistic biochar molecular simulations\cite{wood2024developing,wood2024developingb,ngambia2024development,wood2026unravelling} to quantify how (i) pyrolysis temperature and (ii) feedstock-dependent biochar chemical structures control Mn(II) association with the carbonaceous material. Crucially, the simulations provide an ash-free baseline and atomistic level detail of the adsorption mechanism that cannot be resolved from bulk experiments alone.

Furthermore, we complement this work with targeted fixed-bed column and batch experiments using oilseed rape straw biochars, identified by Pulcher \emph{et al.} as consistently high-performing for Mn removal.\cite{roberta2026}
We extend sampling across the pyrolysis temperature range by including 350\textdegree C, alongside standard 550\textdegree C and 700\textdegree C, while explicitly monitoring pH evolution and cation release. 
Integrating these datasets enables mechanistic attribution of observed Mn removal to precipitation-led versus surface-complexation-led pathways, and provides transferable chemical design criteria for Mn-targeted biochars.

%%%%%%%%%%%%%%%%%%%%%%%%%%%%%%%%%%%%%%%%%%%%%%%%%%%%%%%%%%%%%%%%%%%%%
%                 MATERIALS AND METHODS
%%%%%%%%%%%%%%%%%%%%%%%%%%%%%%%%%%%%%%%%%%%%%%%%%%%%%%%%%%%%%%%%%%%%%

\section{Materials and Methods}

\subsection {Molecular models and simulations}

\subsubsection{Choice of key descriptors for the development of biochar molecular models}

Our selection of biochars is informed by the study of Pulcher \emph{et al.},\cite{roberta2026} where the performance of a set of Standard Biochars by the UK Biochar Research Center (UKBRC) was studied. Woody biochars (willow chips and softwood derived) generally had significantly lower Mn removal capacity than those derived from straw (oilseed rape straw and wheat straw) -- therefore, we compare their experimentally measured characteristics, Table \ref{SI-tab:BC_exp_prop}, SI. 
The biochars reported to have good performance (oilseed rape straw derived: OSR550, OSR700; and wheat straw derived: WS550, WS700) have a higher ash content ($\sim$20\%), associated with a higher pH ($\sim$9-10), and also feature a higher amount of nitrogen than those derived from softwood (SW550 and SW700). 

Since our modelling goal is to decouple interactions between Mn and biochar molecular structure from ash effects, we use molecular models that represent only the carbonaceous fraction and surface functionalities.
To capture how biochar-forming structures produced at high and low temperatures control Mn adsorption, we built models representative of  400\textdegree C and 800\textdegree C pyrolysis; these temperatures bracket the main changes in functionality. 
To sample feedstock effects, we developed four models spanning two feedstocks (wood and straw) and these two pyrolysis temperatures. Following the experimental naming convention (W = wood, S = straw; last three digits = temperature in \textdegree C), the models are denoted W400 and W800 for woody biochar, and S400 and S800 for straw-derived biochar produced at 400\textdegree C and 800\textdegree C, respectively.

A key requirement for representativeness is chemical composition, i.e., the presence and type of functional groups. Low-temperature biochars retain a variety of oxygenated functionalities, including aromatic \ce{-OH}, \ce{C-OCH3}, \ce{C-O-C}, and other oxygen functional groups.\cite{wood2024developing,wood2024developingb} 
For straw biochars with measurable nitrogen content, this also includes pyridinic-N and pyrrolic-N groups.\cite{yuan2018migration} 
High-temperature biochars contain only stabilised oxygen groups (\ce{C-O-C} and \ce{C=O}) and, for straw-derived biochars, nitrogen groups, such as quaternary-N.\cite{wood2024developingb,yuan2018migration}

To generate the biochar models, we follow our previously developed methodology; see Wood \emph{et al.} and Ngambia \emph{et al.} for full details.\cite{wood2024developing,ngambia2024development} 
Briefly, we first identify the set of descriptors that define the biochar of interest (Table \ref{SI-tab:target_BC}, SI); 
we then construct molecular building blocks (Figure \ref{fig:Biochar_blocks}, SI) representative of chemical descriptors of these biochar. These blocks are assembled to form a solid biochar matrix, validating the emerging physicochemical properties against the set targets. The final biochar structures are shown in Figure \ref{fig:Biochar_models} (SI), and their characterisation, including the surface-exposed group density, is given in Table \ref{tab:model_BC} (SI).

To study biochar systems at pH above their point of zero charge (PZC), we created surface-deprotonated models. For these systems, we randomly deprotonate half of the surface-exposed functional groups that can be deprotonated. 
Accounting for the functional groups present, deprotonation is only possible for the low-temperature biochars that contain \ce{-OH} groups. The deprotonated models are thereafter identified with `-DP' suffix, i.e., W400-DP and S400-DP.

%#########################################################################
\subsubsection{Surface-exposed biochar model construction and simulation}

Based on target chemical descriptors (H/C, O/C, N/C, aromaticity, and functional-group distributions; Table \ref{SI-tab:target_BC}, SI), molecular building blocks were constructed using Marvin Sketch 24.1.3\cite{Marvin2021} and parameterised with OPLS-AA force field, assisted by PolyParGen.\cite{jorgensen1996development,yabe2019development} 
Following our established workflow,\cite{wood2024developing,ngambia2024development} representative ensembles of blocks were assembled and condensed into bulk biochar via step-wise annealing, validated against experimental physicochemical targets, and converted to solvated biochar slabs by expanding the simulation cell normal to the surface to create a vacuum, which is then filled with SPC water.\cite{berendsen1981interaction}
\ce{Mn^2+} and \ce{Cl-} ions were added to the aqueous phase; force field parameters were taken from Li \emph{et al.}\cite{li2017metal} 
To enable comparison across models with different exposed surface areas, Mn loading was held constant per surface area (system compositions in Table \ref{tab:system_setup}, SI). 
Deprotonated variants were generated by deprotonating 50\% of water-accessible hydroxyl groups and adjusting \ce{Cl-} ions to maintain charge neutrality.

Molecular dynamics simulations were performed with GROMACS 2022.4.\cite{bauer2022gromacs} Periodic boundary conditions were applied in all directions. 
Systems were energy-minimised and equilibrated prior to 50\,ns production runs at 300 K and 1 bar. Trajectory convergence was assessed by RMSD, and the final 15 ns were used for analysis. 

Mn association with the biochar surface was analysed using linear density profiles and Mn--heteroatom radial distribution functions (RDF). Inner-sphere complexation was defined using a 0.30 nm Mn--O/N cutoff, while outer-sphere association was defined by Mn within 0.60 nm of surface heteroatoms, with inner-sphere contributions excluded. These cut-off distances are based on RDF.

A step-by-step description of model construction, validation, solvation/ion loading, and surface deprotonation, full molecular dynamics simulation protocols and analysis details, and access to molecular structures and models used are provided in Section \ref{SI:detailed_model_construction}, SI.

%%%%%%%%%%%%%%%%%%%%%%%%%%%%%%%%%%%%%%%%%%%%%%%%%%%%%%%%%%%%%%%%%%%%%%

\subsection {Experimental studies of Mn removal by biochar materials}

Oilseed rape straw biochars produced at the UKBRC were investigated: two standard biochars produced at 550\textdegree C (OSR550) and 700\textdegree C (OSR700),\cite{mavsek2018consistency} and an additional oilseed rape biochar produced at 350\textdegree C following the UKBRC standard production protocol (OSR350). Batch and fixed-bed column experiments were performed at an initial solution pH of 4 to quantify Mn removal and associated changes in solution chemistry (pH and released cations).

Batch experiments (triplicate) were conducted under shaking for 24 h, followed by syringe filtration and analysis of filtrates by ICP-OES. Fixed-bed column experiments were operated in recirculation mode for 300 min with time-resolved sampling of effluents for pH and ICP-OES analysis; post-exposure biochars were recovered and characterised by FTIR. Control experiments quantified Mn precipitation from biochar-free solutions as a function of pH, and \ce{N2} BET surface area was measured for OSR350. Full experimental procedures and instrumental parameters are provided in Section~\ref{SI-sec:exp_methods}, SI.

%%%%%%%%%%%%%%%%%%%%%%%%%%%%%%%%%%%%%%%%%%%%%%%%%%%%%%%%%%%%%%%%%%%%%
%             RESULTS    &     DISCISSION
%%%%%%%%%%%%%%%%%%%%%%%%%%%%%%%%%%%%%%%%%%%%%%%%%%%%%%%%%%%%%%%%%%%%%

\section{Results and Discussion}

The aim of this work is to examine each possible mechanism for Mn sequestration from the solution assisted by biochar: (i) precipitation of oxidized Mn species from the solution at high pH, (ii) exchange of cations within biochar for Mn from the solution, and (iii) role of biochar surface-exposed functional groups and importance of its molecular composition. To examine each of these mechanisms independently, we use experimental studies and molecular dynamics simulations synergistically. 

In the laboratory, we carry out a combination of batch and column experiments. 
Column studies provide information on the uptake of Mn by biochar over time, while monitoring the change in the solution pH. 
Batch studies allow us to compare how the presence of Mn impacts the cations leached from the biochar into water. 
Molecular dynamics simulations are ideal for studying a pure biochar system, removing any ash contributions, focusing on the specific biochar surface functional groups and examining the interactions at the atomic level.
For molecular modelling studies, we examine Mn adsorption by biochars produced from wood and straw materials at low (400\textdegree C) and high (800\textdegree C) pyrolysis temperature. These systems are fully representative of their experimental counterpart chemical compositions, but are completely free from any inorganics, i.e., ash.

%#########################################################################
\subsection{Does elevated pH drive precipitation as the key mechanism for aqueous \ce{Mn^2+} sequestration?}

\begin{figure}[ht!]
    \centering
    \includegraphics[width=0.9\linewidth]{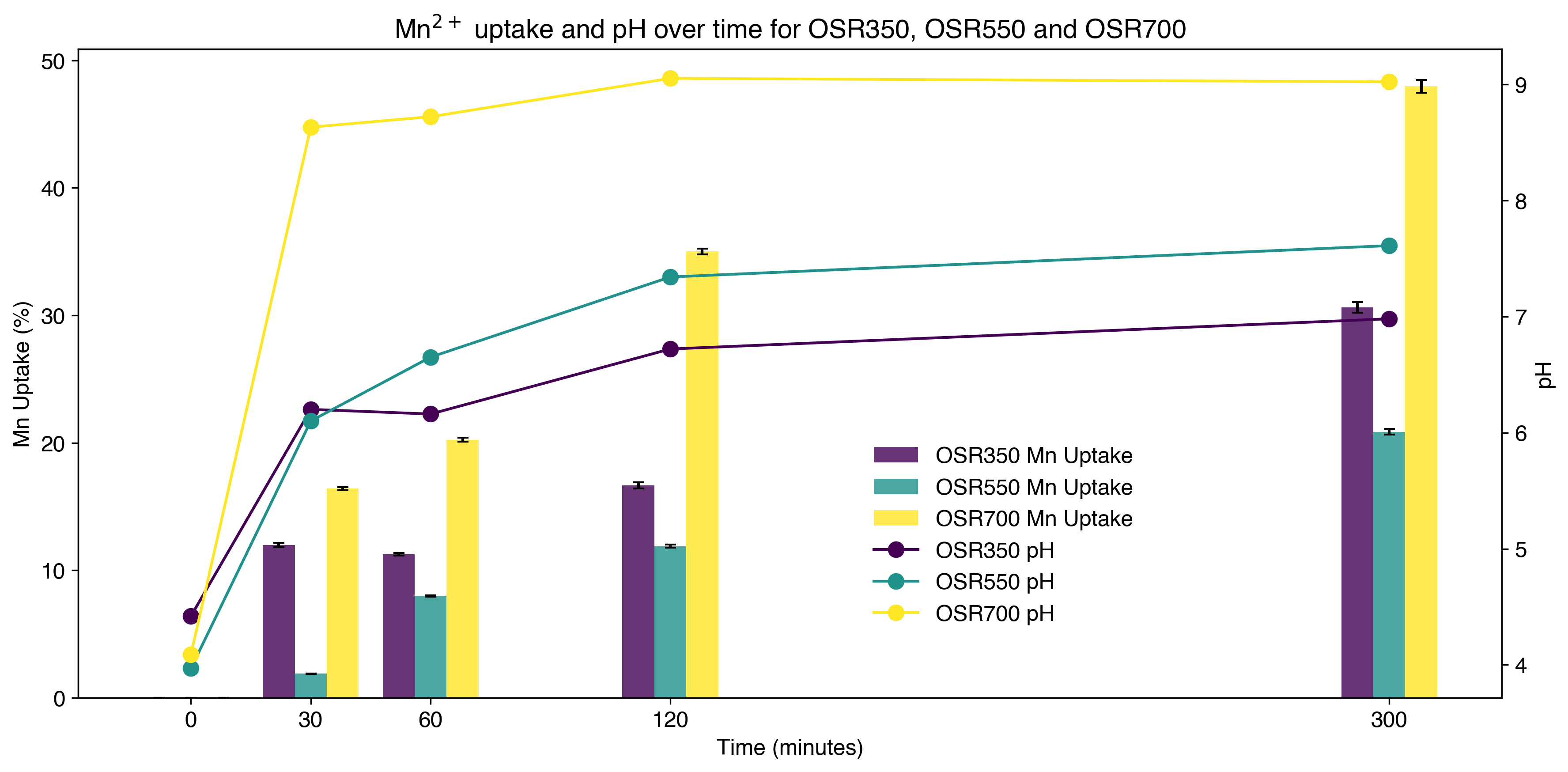}
    \caption{Manganese uptake over time from 5 ppm solution by OSR350 (purple), OSR550 (cyan) and OSR700 (yellow), and the changes in pH of the solution.}
    \label{fig:removal_pH}
\end{figure}

In the column studies, the removal of Mn from the solution over time was monitored alongside the solution pH (Figure \ref{fig:removal_pH}). 
Over 300 min, both Mn removal and solution pH increased progressively. For OSR700, pH rose rapidly from 4, stabilising at 9 within 90 min, and this biochar achieved the greatest removal of $\sim$50\%.
This behaviour is consistent with reported thresholds for Mn(II) oxidation and precipitation, which commonly occur above pH 8-9, depending on the dissolved oxygen and catalytic surfaces present
(Figure \ref{fig:removal_pH_pure}, SI).\cite{aziz1992influence} 

Interestingly, the biochar produced at lower temperatures -- OSR350 and OSR550 -- still showed a good Mn removal capacity (20-30\%), while the pH of the solution only increased from 4 to 7-7.5. 
We note that these removal capacities are reported per mass of biochar, whereas surface area is often taken as a more intrinsic descriptor, even if \ce{N2} BET is not truly representative of water-accessible area.
OSR350 has a very low BET surface area (2.1 m$^2$ g$^{-1}$) compared to OSR700 (25.2 m$^2$ g$^{-1}$), yet still achieves Mn removals only modestly lower than OSR700 (30\% vs 48\%). (\ce{N2} BET of OSR350 measured in this work, see Figure \ref{fig:BET}, SI; for others see Table \ref{SI-tab:BC_exp_prop}, SI). 
Normalised per surface area, this corresponds to $\sim$0.15 mg m$^{-2}$ for OSR350 and $\sim$0.02 mg m$^{-2}$ for OSR700 -- eight times lower, despite OSR700 achieving comparable bulk removal.
This suggests that surface chemistry, rather than geometric surface area alone, may govern Mn uptake on these materials.

Overall, these results indicate that the removal of Mn may be driven by some other mechanisms besides precipitation in a high-pH environment produced by biochar.
Interestingly, OSR700 and OSR550 have similar ash content (22.0 vs. 19.5 wt\%; Table \ref{SI-tab:BC_exp_prop}, SI), so pH-elevating cation release may reflect both ash dissolution and cationic exchange -- a distinction investigated in the following section.

%#########################################################################
\subsection{Is cationic exchange the dominant mechanism for \ce{Mn^2+} uptake by biochars?}

Oilseed rape straw biochars exhibit high cation exchange capacity (CEC; 856 mmol kg$^{-1}$ for OSR550 and 735 mmol kg$^{-1}$ for OSR700; Table \ref{SI-tab:BC_exp_prop}, SI), indicating substantial capacity for charge-compensating uptake of dissolved metal cations.
To investigate the extent to which cation exchange contributes to \ce{Mn^2+} removal, we measured the leached exchangeable cations from biochar, in particular focusing on alkali and alkaline-earth cations, as their release would also raise the pH of the solution.

Table \ref{tab:cations} summarises dissolved \ce{K+}, \ce{Ca^2+} and \ce{Mg^2+} concentrations after contacting the biochars with solutions in the absence and presence of \ce{Mn^2+} (at 4.38 ppm). 
We omit \ce{Na+} because \ce{NaOH} was used for pH control; consequently, measured \ce{Na+} concentrations cannot be attributed to biochar-driven processes. Furthermore, \ce{Na+} precludes strict charge-balance calculations, and the trends in released cations are therefore interpreted qualitatively.

In all systems, \ce{K+} is the dominant released cation, consistent with OSR biochars containing a large, readily mobilisable K pool. Importantly, addition of \ce{Mn^{2+}} systematically increased \ce{K+} release across all biochars (Table \ref{tab:cations}). 

For OSR350, \ce{Mn^{2+}} removal was lower than for the higher-temperatures biochars and coincided with higher mean concentrations of all measured indigenous cations in solution (\ce{K+}, \ce{Ca^{2+}} and \ce{Mg^{2+}}), consistent with Mn uptake being coupled to displacement of a mixed pool of counter-ions from heterogeneous exchange sites. 
By contrast, Mn removal by the higher-temperature biochars was near-absolute (OSR550 at 97\%, OSR700 at 98\%), but the accompanying cation response was dominated by enhanced \ce{K+} mobilisation, while dissolved \ce{Ca^{2+}} and \ce{Mg^{2+}} were suppressed relative to Mn-free controls. 
The pronounced increase in \ce{K+} release for OSR550 (near 40\%), therefore, points to monovalent cations (primarily \ce{K+}, and potentially \ce{Na+}) as the main exchange partners under these conditions, whereas the limited divalent-cation release implies that \ce{Ca^{2+}}/\ce{Mg^{2+}} are not exchanged stoichiometrically and/or are retained in sparingly soluble mineral forms. 
For OSR700, the smaller (and less certain) increase in \ce{K+} (+20\%), despite comparable Mn removal, suggests an increased relative contribution of non-exchange pathways (e.g., Mn surface precipitation and/or association with ash/mineral domains) alongside any ion-exchange component. Overall, the non-stoichiometric relationship between Mn removed and cations released indicates that cation exchange contributes to Mn uptake, most clearly for OSR350 and OSR550, but does not solely account for Mn immobilisation in the high-temperature biochars.

To contextualise the magnitude of cation release, we compared dissolved concentrations with total elemental content, reported for the standard biochars (Table \ref{SI-tab:BC_exp_prop}, SI). UKBRC values were obtained by Aqua Regia digestion (total element), whereas the present measurements reflect only the dissolved (leached/exchanged) fraction; total contents therefore provide an upper bound on possible solution concentrations.
For OSR550 and OSR700, total K contents of 28,600 and 29,800 mg kg$^{-1}$, respectively, correspond to 57 and 59 ppm at a biochar loading of 2 g L$^{-1}$. The observed K concentrations are of the same order (Table \ref{tab:cations}), indicating that a substantial fraction of K is readily mobilised and available for exchange/displacement during Mn uptake. This is also in agreement with values reported in other works, where OSR biochars are used for removal of divalent heavy metals.\cite{lam2019removal}
By contrast, despite high total Ca contents for OSR550/OSR700 (22,000 and 23,800 mg kg$^{-1}$, respectively), dissolved Ca remained low. This discrepancy is consistent with Ca being predominantly present as \ce{CaCO3}, whose solubility decreases under alkaline conditions, thereby limiting its presence in the aqueous phase.
Moreover, such carbonate-bearing mineral phases can provide a carbonate source and nucleation surfaces that favour Mn carbonate/hydroxide precipitation, offering a plausible explanation for the near-quantitative Mn removal by OSR550/OSR700 despite limited divalent-cation release.

\begin{table}
    \centering
    \begin{tabular}{c|ccc|c}
    \hline
        System /Cation (ppm)                & \ce{K+}  & \ce{Ca^2+}  & \ce{Mg^2+}  & \ce{Mn^2+}  \\
        \hline
         OSR350 w/o \ce{Mn^2+} & 67.0±4.5 & 21.6±3.1 & 6.3±0.8 & 0.008±0.008\\
         OSR550 w/o \ce{Mn^2+} & 48.2±5.6  & 4.2±0.2 & 0.78±0.04 & 0.003±0.001\\
         OSR700 w/o \ce{Mn^2+} & 56.0±4.2 & 2.5±0.1 & 0.44±0.02 & 0.002±0.001\\
         \hline
         OSR350 with \ce{Mn^2+} & 80.0±10.9  & 25.7±6.8 & 8.1±2.3 & 1.21±0.33\\
         OSR550 with \ce{Mn^2+} & 65.7±0.65 & 2.5±0.3 & 0.46±0.1 & 0.12±0.01 \\
         OSR700 with \ce{Mn^2+} & 67.3±7.5 & 1.83±0.3 & 0.40±0.01 & 0.09±0.05\\
         \hline
         OSR350 diff./ \ce{Mn^2+} removal & 13.0 (+19\%)  & 4.0 (+19\%) & 1.8 (+29\%) & * 3.16 (72\%) \\
         OSR550 diff./ \ce{Mn^2+} removal & 17.6 (+37\%) & -1.7 (-41\%) & -0.3 (-41\%) & * 4.26 (97\%) \\
         OSR700 diff./ \ce{Mn^2+} removal & 11.4 (+20\%) & -0.65 (-26\%) & -0.04 (-10\%) & * 4.28 (98\%) \\   
        \hline
    \end{tabular}
    \caption{Alkali and alkaline-earth metals leached from biochars without manganese (top set) and with 4.38 ppm \ce{Mn^2+} (middle set). The difference between the two system is given as the bottom set, including percentage increase in brackets. *for \ce{Mn^{2+}} removed amount and percentage are shown.} 
    \label{tab:cations}
\end{table}

To further probe the mode of Mn retention, FTIR spectra of the pristine biochars and the Mn-exposed biochars were compared (Figure \ref{fig:FTIR}, SI). 
The pristine OSR700 spectrum already exhibits a pronounced band at $\sim$425 cm$^{-1}$, which is commonly assigned to metal–O lattice vibrations associated with the mineral/ash fraction. In the context of Mn, this would be associated with octahedral-like sites and found in crystalline \ce{Mn3O4} or \ce{Mn2O3} phases. As a result, any additional contribution arising after Mn exposure are superimposed on this pre-existing feature and, even though noticeable by eye, are not confidently distinguishable by FTIR. Taken together, the limited spectral evolution is consistent with Mn immobilisation occurring predominantly via inorganic pathways, in agreement with the limited release of divalent base cations in the corresponding leachates.

OSR350 and OSR550, in contrast, exhibit clear spectral changes after exposure to \ce{Mn^{2+}}, including altered band intensities/positions associated with oxygen-containing functionalities and the emergence of a broad feature in the 700-400 cm$^{-1}$ region. These low-wavenumber broad bands are characteristic of \ce{Mn-O} bound to organic ligands and experiencing multiple environments (such as chelated and bridging).\cite{hu2024simultaneous,yang2022solid} 
Furthermore, a slight shift (by under 50 cm$^{-1}$) in the carbonyl band position suggests the emergence of bridging or bidentate chelation.
While FTIR alone cannot unequivocally distinguish inner-sphere surface complexation from the formation of fine Mn (oxy)hydroxide or surface-precipitated phases, the observed spectral perturbations for OSR350/OSR550 indicate a larger contribution from surface-associated Mn–O interactions in these lower-temperature biochars. 
In combination with the cation release trends, these results support a mixed mechanism in which cation exchange contributes to Mn uptake and is complemented by Mn–O surface interactions, whereas OSR700 uptake is more consistent with mineral-controlled retention (precipitation/co-precipitation) with limited involvement of organic functional groups detectable by FTIR.

%#########################################################################
\subsection{What role do surface functional groups play in \ce{Mn^{2+}} complexation?}

To isolate the role of surface chemistry in \ce{Mn^{2+}} uptake, we performed molecular dynamics simulations for four biochar models spanning two feedstocks (wood and straw) and two pyrolysis temperatures (400\textdegree C and 800\textdegree C). Because only the low-temperature models contain deprotonatable --OH groups, we additionally constructed partially deprotonated variants (W400-DP and S400-DP) to represent conditions above the biochar PZC, where negatively charged surface sites are available.

\begin{figure}
    \centering
    \includegraphics[width=1\linewidth]{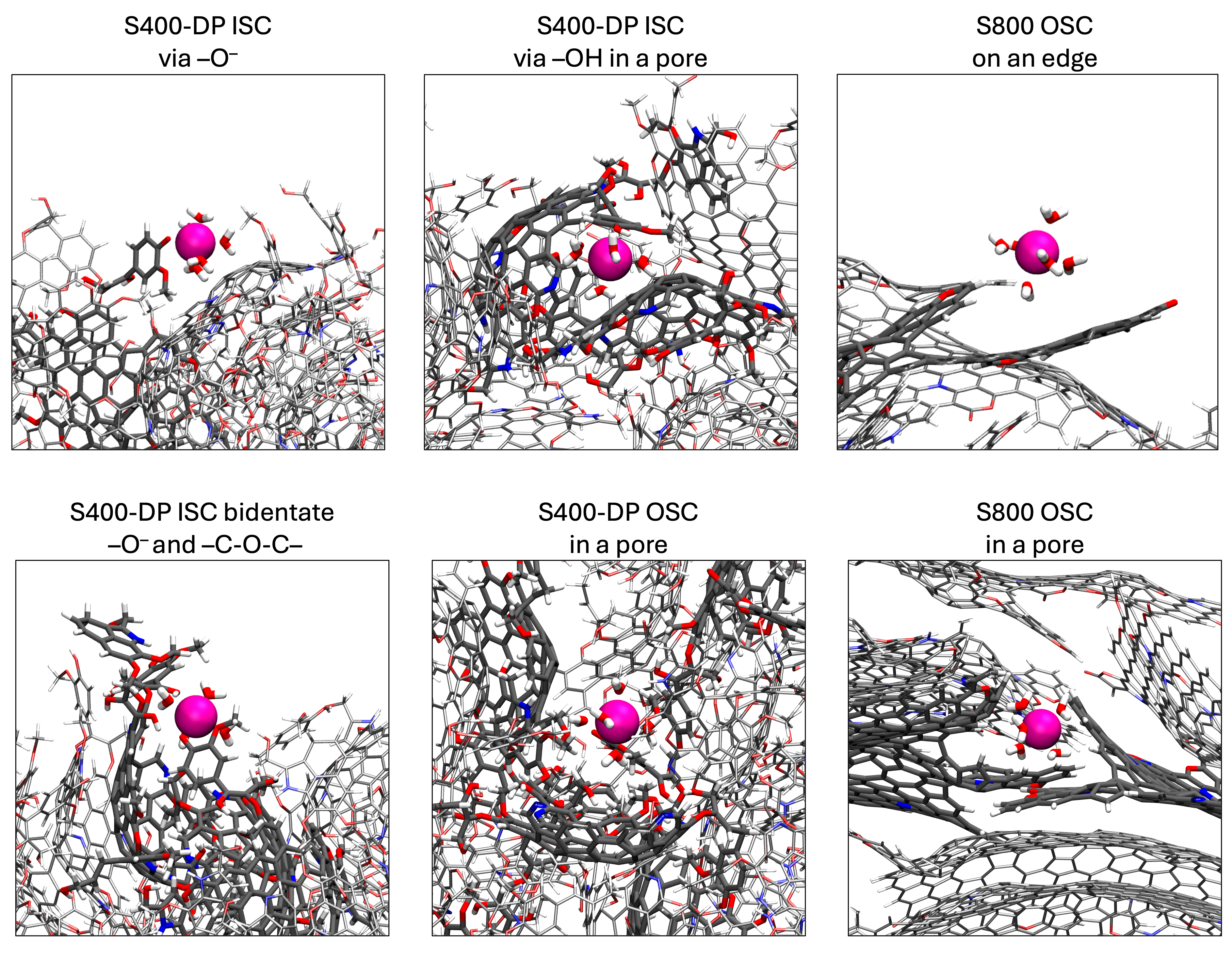}
    \caption{Representative Mn adsorption motifs on biochar, illustrated for S400-DP and S800. In the deprotonated systems, the dominant motif is an inner-sphere complex (ISC) formed by monodentate coordination to \ce{-O^-}. Less frequent ISC configurations include bidentate binding to \ce{-O^-} and a neighbouring \ce{-C-O-C-} group, as well as inter-pore adsorption to a neutral \ce{-OH} group. Outer-sphere complexes (OSCs) are observed within pores in all biochar systems. Nearby biochar are shown as thick licorice representations, more distant moieties are rendered as semitransparent. Colours: C -- gray, O -- red, N -- blue, H -- white. \ce{Mn^{2+}} are shown as magenta van der Waals spheres. Only first-shell water molecules are shown in licorice representation (O -- red, H -- white); bulk water is omitted for clarity.}
    \label{fig:mechanisms}
\end{figure}

Representative equilibrated systems are shown in Figure \ref{fig:rendering_biochar_Mn_system} (SI). In the protonated low-temperature models (W400 and S400), linear density profiles show no enrichment of \ce{Mn^{2+}} at the interface relative to the bulk solution (Figure \ref{fig:normalised_density}, SI), similarly RDF analysis shows no significant association between \ce{Mn^{2+}} and surface functional groups (Figure \ref{fig:RDF}, SI).
The high-temperature models (W800 and S800) show the same absence of interfacial enrichment and inner-sphere coordination. However, their greater porosity allows water and ions to enter the pore network, giving a non-zero probability of \ce{Mn^{2+}} within the biochar matrix. Together with the low density of surface functional groups in W800 and S800 (Table \ref{tab:model_BC}), this indicates that high-temperature biochars mainly provide pore volume for hydrated \ce{Mn^{2+}}, rather than sites for strong inner-sphere complexation. This mechanism is illustrated in Figure \ref{fig:mechanisms}.
The modest Mn uptake observed for these systems (Table \ref{tab:Mn_adsorption_sphres}), therefore, arises from weak outer-sphere association and pore diffusion, implying that the higher experimental Mn removal by high-temperature OSR biochars is more likely dominated by ash-driven pH increase and subsequent Mn(III/IV) (hydr)oxide precipitation than by direct adsorption to the carbonaceous surface.

This behaviour changes markedly in the deprotonated models (W400-DP and S400-DP), where density profiles show clear \ce{Mn^{2+}} accumulation at the surface (Figure \ref{fig:normalised_density}, SI), and RDFs confirm specific interactions with deprotonated groups (Figure \ref{fig:deprotonated_RDF}, SI). The strongest correlations are observed for phenolic/hydroxyl oxygen at $\sim$0.2 nm and anisole oxygen at $\sim$0.25 nm, consistent with inner-sphere coordination. 
By contrast, the Mn–pyran oxygen distance ($\sim$0.45 nm) is consistent with outer-sphere association via an intervening water molecule, as is the distance to pyridine nitrgen ($\sim$0.45 nm) and pyrrole nitrogen lies further still ($\sim$0.55 nm).\cite{persson2024structure}
Representative adsorption motifs are shown in Figure \ref{fig:mechanisms}.

\ce{Mn^{2+}} therefore forms both inner- and outer-sphere complexes at the biochar surface, with the quantitative contributions summarised in Table \ref{tab:Mn_adsorption_sphres}. 

Inner-sphere complexation is observed only for the deprotonated systems, accounting for 48\% of Mn uptake in W400-DP and 57\% in S400-DP. Although both models have similar SASA ($\sim$3 nm$^2$ nm$^{-2}$), S400 exposes 15\% more oxygen-containing groups per SASA than W400 and 18\% more total surface groups owing to the additional nitrogen functionalities (Table \ref{tab:system_setup}, SI), confirming that deprotonated surface group density, rather than surface area, is the primary control on Mn complexation.

Outer-sphere association is present across all systems and reflects two distinct contributions. In the deprotonated models it is slightly enhanced relative to the protonated counterparts (W400: 10\% vs.\ 12\%; S400: 5\% vs.\ 8\%), consistent with negative surface charge favouring retention of hydrated cationic species in addition to direct inner-sphere binding. 
In the high-temperature models (W800, S800), where inner-sphere sites are absent, outer-sphere association (13-14\%) is instead driven by pore diffusion: the greater microporosity of these biochars allows hydrated \ce{Mn^{2+}.6H2O} (diameter $\sim$0.85 nm, Figure \ref{fig:hydrated_Mn_RDF}, SI) to enter pore regions above $\sim$1 nm, which is also seen directly in the density profiles (Figure \ref{fig:normalised_density}, SI).

\begin{table}
\centering
\resizebox{\columnwidth}{!}{%
\begin{tabular}{c|ccc|ccc}
\hline
\multirow{2}{*}{Model Name} & 
    \multicolumn{3}{c|}{Inner sphere \ce{Mn^{2+}} adsorption} & 
    \multicolumn{3}{c}{Outer sphere \ce{Mn^{2+}} adsorption} \\
\cline{2-7}
& \makecell{µmol\\ m$^{-2}$} & 
\makecell{µmol\\ per SASA\\ (m$^{-2}$)} & 
\% removed & 
\makecell{µmol\\ m$^{-2}$} & 
\makecell{µmol\\ per SASA\\ (m$^{-2}$)} & 
\% removed \\
\hline
Woody400       & 0 & 0 & 0 &    0.06$\pm$0.03    & 0.02$\pm$0.01 &  9.98$\pm$0.05  \\
Woody800       & 0 & 0 & 0 &    0.14$\pm$0.02    & 0.02$\pm$0.003 & 13.56$\pm$0.02  \\
\hline
Straw400       & 0 & 0 & 0 &    0.03$\pm$0.01    & 0.01$\pm$0.004 & 4.52$\pm$0.02   \\
Straw800       & 0 & 0 & 0 &    0.13$\pm$0.02    & 0.02$\pm$0.003 & 12.63$\pm$0.02 \\
\hline
Woody400-DP    & 0.37$\pm$0.03 & 0.10$\pm$0.003 & 47.68$\pm$0.03 & 0.07$\pm$0.02 & 0.02$\pm$0.007 & 11.57$\pm$0.01  \\
Straw400-DP    & 0.41$\pm$0.01 & 0.12$\pm$0.003 & 57.35$\pm$0.02 & 0.05$\pm$0.01 & 0.02$\pm$0.003 & 7.85$\pm$0.02 \\
\hline
\end{tabular}
}
\caption{Summary of Mn inner-sphere ($<$0.3 nm) and outer-sphere (0.3-0.6 nm) complexation on biochar. Values are reported per projected $xy$-area, per SASA, and as the percentage of dissolved Mn removed by each mechanism.}
\label{tab:Mn_adsorption_sphres}
\end{table}

%#########################################################################
\subsection{A multi-step mechanism for manganese sequestration by biochar}

\begin{figure}[ht!]
    \centering
    \includegraphics[width=1\linewidth]{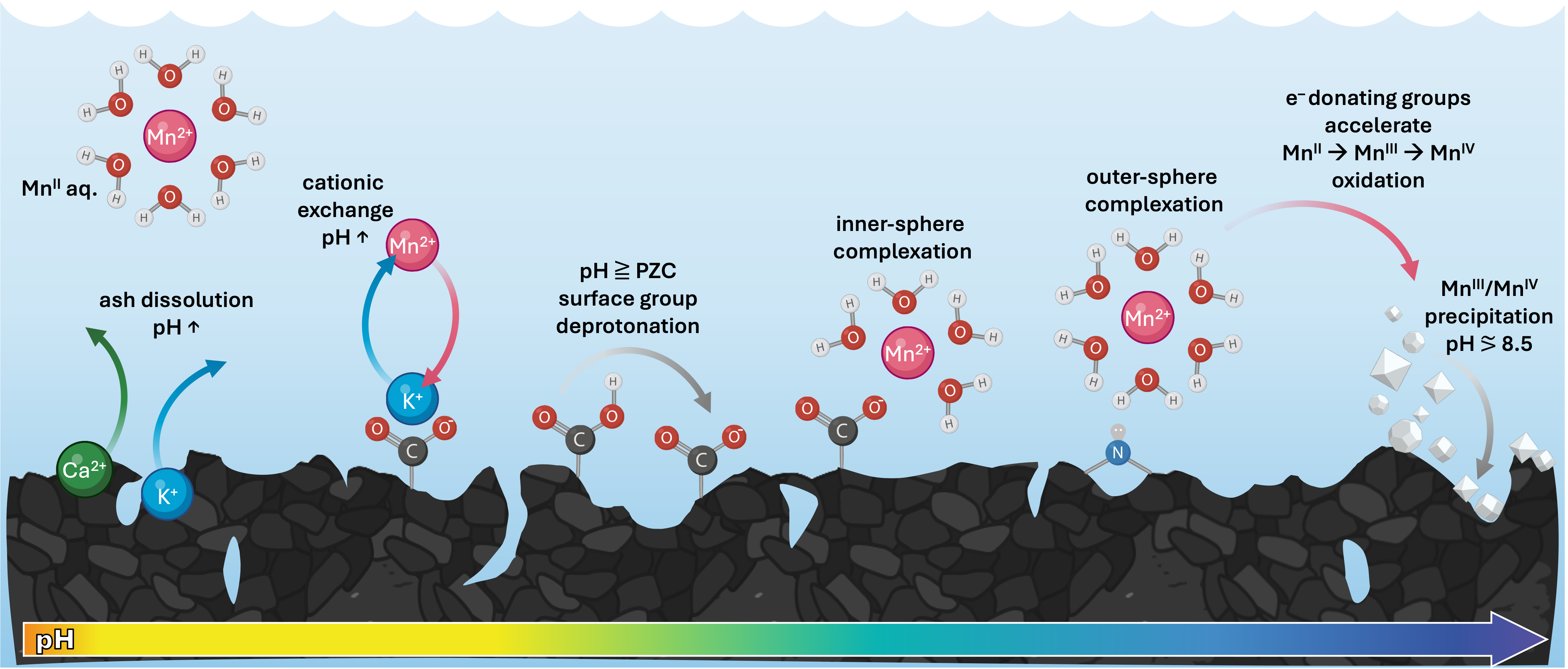}
    %\caption{Mechanistic schematic of aqueous Mn(II) sequestration by biochar. Upon contact, (1) rapid ash dissolution and cation exchange release basic cations and elevate pH; (2) rising pH drives deprotonation of surface functional groups and builds negative surface charge; (3) deprotonated sites bind \ce{Mn^2+} via inner‑sphere complexes, while electron-donating sites form outer-shell complex with hydrated \ce{Mn^2+}; and (4) at sufficiently high pH (further speed up by electron-donating groups) Mn(II) oxidizes to Mn(III/IV) and precipitates as (hydr)oxide; overall \ce{Mn^2+} removal is a coupled exchange $\rightarrow$ complexation $\rightarrow$ precipitation sequence whose relative importance depends on feedstock and pyrolysis temperature of biochar production.}
    \caption{Multi-step mechanism for aqueous Mn(II) sequestration by biochar: 
    (1) cation exchange and ash dissolution raise solution pH; 
    (2) pH increase drives surface group deprotonation; 
    (3) deprotonated O/N sites bind \ce{Mn^{2+}} via inner- and outer-sphere complexation; 
    (4) at pH~$\gtrsim$8.5, Mn(II) oxidises and precipitates as Mn(III/IV) (hydr)oxides. 
    Relative contributions of each step depend on biochar feedstock and pyrolysis temperature.}
    \label{fig:SCHEMA}
\end{figure}

Overall, our experimental and modelling results support a multi-step mechanism for Mn removal by biochar.
Upon initial contact of biochar with Mn solution at low pH, readily exchangeable cations (predominantly monovalent cations, but also with smaller contributions from divalent ones) are released from the biochar (Table \ref{tab:cations}). 
This reflects cation exchange between solution-phase \ce{Mn^{2+}} and monovalent/divalent cations associated with the biochar surface and mineral phases. 
The cation stoichiometry (Table \ref{tab:cations}) points to preferential \ce{Mn^{2+}}/\ce{K+} exchange, especialy at higher pyrolysis temperatures.

The cumulative release of basic cations raises the solution pH towards, and often beyond, the point of zero charge (PZC) of the biochar (Figure \ref{fig:removal_pH}, Table \ref{SI-tab:BC_exp_prop}, SI). 
When pH$>$PZC, surface oxygen-containing functional groups partially deprotonate, generating negatively charged sites. 
Deprotonated models confirm pronounced \ce{Mn^{2+}} surface enrichment, with $\sim$50\% removed via inner-sphere complexation (Table \ref{tab:Mn_adsorption_sphres}).
In parallel, the pH increase above 8-8.5 promotes Mn(II) oxidation and precipitation of Mn(III/IV) (hydr)oxides, a pathway shown to be negligible below pH 8 and dominant for OSR700 (Figure \ref{fig:removal_pH_pure}, SI).

The presence of electron-rich nitrogen functionalities further modulates Mn binding. In the straw-derived models, pyridinic and pyrrolic N sites form outer-sphere coordination complexes with \ce{Mn^{2+}} at distances consistent with a retained hydration shell, in line with known the behaviour of \ce{Mn^{2+}-N} ligands.\cite{uzal2020mn2+}
These N-donor sites are expected to stabilise Mn surface complexes and potentially speed-up Mn oxidation kinetics, as suggested by studies of Mn redox chemistry in the presence of multidentate ligands.\cite{morgan2021rates,rosso2002outer,morgan2005kinetics,luther2005manganese} 
Although our classical molecular dynamics simulations cannot capture redox transformations explicitly, the combination of FTIR evidence for Mn-O bonding on low-temperature OSR biochars (Figure \ref{fig:FTIR}, SI), high cation exchange capacity, and strong modelled Mn–O/N interactions, together, point to a coupled exchange-complexation pathway that precedes, and likely accelerates, heterogeneous Mn oxidation.

In natural environments, Mn(III) is increasingly recognised to occur predominantly as soluble Mn(III)-ligand complexes, often with humic- or fulvic-like organic matter, rather than as discrete oxide phases.\cite{oldham2019speciation}
This underscores the capacity of electron-rich O- and N-donor ligands to stabilise intermediate Mn oxidation states, and suggests that analogous functionalities on biochar surfaces may similarly bind and modulate the redox speciation of Mn.

We propose a coupled sequence of processes consistent with our experimental and simulation observations, illustrated in Figure \ref{fig:SCHEMA}: 
\begin{enumerate}
    \item  Rapid cation exchange and partial ash dissolution raise solution alkalinity on initial contact with biochar; 
    \item  As pH approaches/exceeds the biochar PZC, surface groups begin to deprotonate, creating negatively-charged binding sites; 
    \item  Deprotonated O‑ and electron‑rich N‑sites promote inner‑sphere and outer‑sphere Mn(II) complexation; 
    \item At sufficiently high pH (8.5-9), further accelerated by complexing to electron-donating functional groups, oxidation of Mn(II) to Mn(III/IV) (hydr)oxides becomes thermodynamically and kinetically favourable, producing particulate Mn phases that dominate removal on ash‑rich biochars.
\end{enumerate}
The relative importance of these steps depends on feedstock, pyrolysis temperature, resulting ash content and surface chemistry, providing clear design levers for tuning biochar towards more efficient and selective Mn removal.

%#########################################################################
\section{Conclusions and Implications for Biochar Design}

In this study, we combined batch and column experiments with molecular dynamics simulations of realistic biochar structures to disentangle the mechanisms by which biochar removes aqueous \ce{Mn^{2+}}. Three coupled processes emerged as central: 
(i) cation exchange and ash dissolution, 
(ii) deprotonation and complexation at surface functional groups, and 
(iii) pH-driven Mn precipitation.

In fixed-bed column experiments, OSR biochars produced at 350-700\textdegree C removed 20-50\% of dissolved Mn from dilute acidic influent (5 mg L$^{-1}$ Mn, pH$_{ini}$ 4, 2 g L$^{-1}$ biochar, 5 h), while batch experiments under equilibrium conditions (4.38 mg L$^{-1}$ Mn, pH$_{ini}$ 4, 5 g L$^{-1}$ biochar, 24 h) yielded substantially higher removals (72-98\%). This difference is consistent with the shorter contact time and lower biochar-to-solution ratio in the column configuration and underscores that the mechanisms identified here operate across both experimental regimes.
In fixed-bed column experiments, high-temperature, ash-rich OSR700 raised pH to 9 and achieved the highest Mn removal, indicating a dominant contribution from Mn(III/IV) (hydr)oxide precipitation. 
In contrast, OSR350 and OSR550 removed less Mn and remained in neutral pH (7-7.5), where precipitation is negligible, yet cation release and FTIR data indicate surface complexation and ion exchange as the operative pathways.

The molecular models show that neutral (protonated) biochar surfaces support only modest outer-sphere \ce{Mn^{2+}} association (5-14\% removal), enhanced by higher porosity in high-temperature materials. Partially deprotonated low-temperature biochars (W400-DP, S400-DP) with negative surface charge and abundant phenolic/OH and N-donor sites, strongly bind \ce{Mn^{2+}}, with $>$50\% of Mn removed via inner-sphere complexes and a further $\sim$10\% via outer-sphere adsorption.
Thus, the density of deprotonated surface groups, rather than surface area alone, is the primary control on Mn complexation by the biochar.
This is corroborated by OSR350, which in fixed-bed studies removed $\sim$30\% of dissolved Mn despite a surface area ten times smaller than OSR700.

These findings show that efficient aqueous Mn sequestration by biochar depends on the interplay between: 
(i) cation exchange capacity and basic cation inventory (controlling pH evolution), 
(ii) the abundance and acidity of oxygenated groups that can deprotonate near environmental pH, and 
(iii) the presence of electron-rich nitrogen functionalities that stabilise Mn surface complexes. 

Therefore, for Mn and other d-metal contaminants, N-enriched, low-to-intermediate temperature biochars with high densities of deprotonatable O-groups appear particularly promising. While their surface areas are smaller than their high-temperature counterparts, the surface area can be further increased by physical methods (e.g., grinding) after pyrolysis, rather than chemical methods, which risk removal of the functional groups.
More broadly, this work illustrates how molecular simulations, anchored to standardised biochar characterisation, can be integrated with targeted experiments to derive mechanistic design rules for sorbents in aquatic pollution control.


\begin{acknowledgement}

The authors thank John Tobin for helping with the BET analysis. AN also acknowledges the University of Edinburgh Doctoral College Scholarship in supporting her PhD research project. All simulations were performed on the Cirrus UK National Tier-2 HPC Service at EPCC (http://www.cirrus.ac.uk) funded by the University of Edinburgh and EPSRC (EP/P020267/1).

\end{acknowledgement}

\begin{suppinfo}

Supporting Information (PDF) includes detailed computational and experimental methodology, biochar characterisation data, model descriptors and compositions, and additional experimental and simulation figures and analyses.

\end{suppinfo}


\bibliography{references}

\clearpage

\includepdf[pages=-]{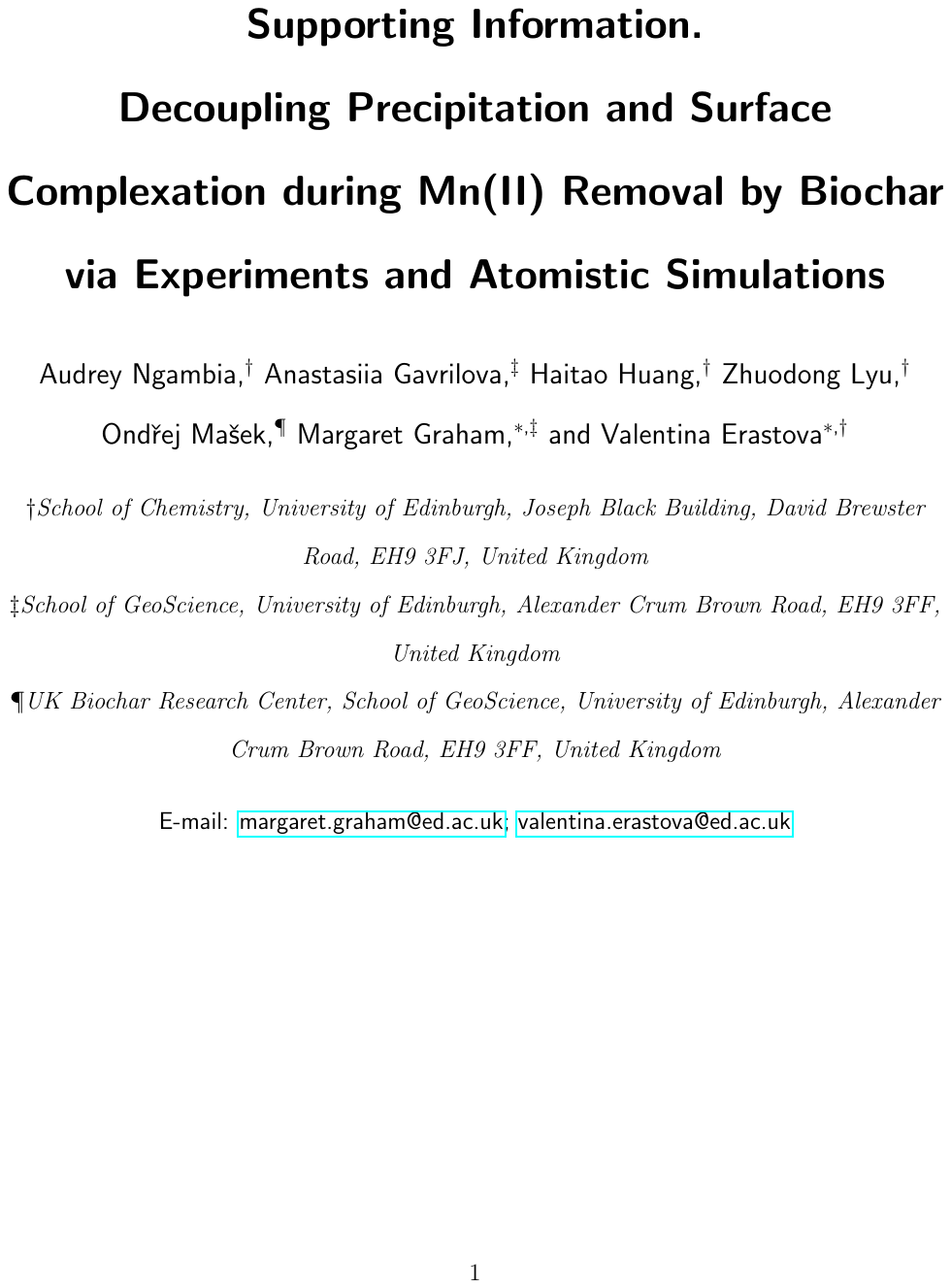}

\end{document}